\documentclass{PoS}

\title{Using software spectrometer to ensure VLBI signal chain reliability}

\ShortTitle{Using software spectrometer to ensure VLBI signal chain reliability}

\author{\speaker{Minttu Uunila} and Juha Kallunki\\
        Aalto University Mets\"ahovi Radio Observatory, Mets\"ahovintie 114, 02540 Kylm\"al\"a, Finland\\
        E-mail: \email{minttu.uunila@aalto.fi}, \email{juha.kallunki@aalto.fi}
}

\author{Guifr\'e Molera Calv\'es\\
        Joint Institute for VLBI in Europe, JIVE, Oude Hoogeveensedijk 4, 7991 PD Dwingeloo, The Netherlands\\
        E-mail: \email{gofrito@gmail.com}
}

\abstract{Software spectrometer (SWSpec) developed for spacecraft tracking can be used to assure VLBI signal chain reliability, and phase stability of a VLBI receiver. Testing performed with SWSpec during pre-operations both saves time, and eases the tests as one does not need to gather, couple and setup the hardware.}

\FullConference{12th European VLBI Network Symposium and Users Meeting\\
                 7-10 October 2014\\
                 Cagliari, Italy}

\begin{document}

\section{Introduction}

\begin{figure}
\begin{center}
  \includegraphics[width=0.65\linewidth]{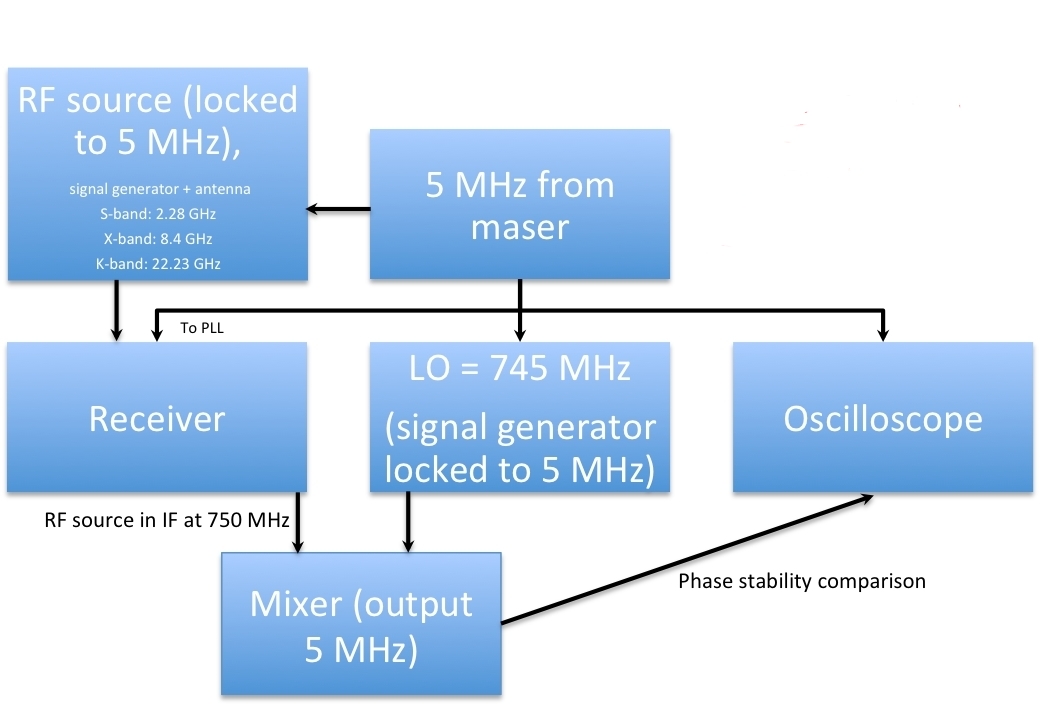}
\end{center}
\caption{Phase stability measurement setup without using a software spectrometer requires several measurement devices.}
\label{fig:coh}
\end{figure}

Software spectrometer (SWSpec, [1]) developed for spacecraft tracking can also be used to assure VLBI signal chain reliability and phase stability of a VLBI receiver. If problems occur during an EVN session, one can use SWSpec without changing the original hardware setup. Testing performed with SWSpec during pre-operations both saves time and eases the tests as one does not need to gather, couple and setup the hardware anymore. In the future testing can be performed by operators that reduces workload of technical staff. 
For the earlier test setup (including a RF source, receiver, 5 MHz from maser, signal generator (LO=745 MHz) locked to 5 MHz, mixer with an output of 5MHz and oscilloscope), see Figure \ref{fig:coh}. 

\section{Software description}

SWSpec is a high performance software spectrometer used to detect spacecraft tones. It accepts VLBA and MkIV formats  and  Mark5A, PC--EVN, Mark5B, VDIF and raw input \cite{molera2013}. All settings are fully adjustable. 

\begin{figure}
\begin{center}
  \includegraphics[width=0.7\linewidth]{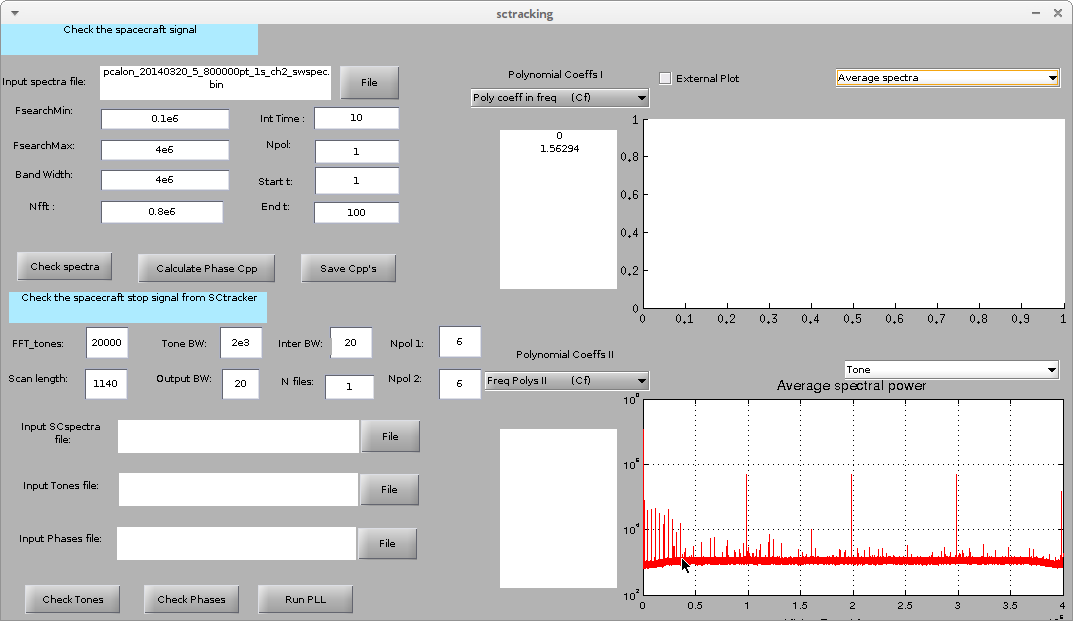}
\end{center}
\caption{sctracking GUI.}
\label{fig:gui}
\end{figure}

SCtrackAnalysis is a Matlab tool with Graphical User Interface (GUI), see Figure \ref{fig:gui}, designed to analyse the results obtained with SWSpec. The scripts can also be used independently as data verification tools. The GUI allows the user to perform the data post-analysis with a single application. All the scripts have been developed and maintained at JIVE by Guifr\'e Molera Calv\'es.

The input parameters for the application are fully configurable and they can be adjusted to the observing and/or processing mode used. Number of FFT points, number of scans, integration time, bandwidth, boundaries around the spacecraft tone (Fsmin and Fsmax) and order of the polynomial fit can be defined by the user \cite{molera2014}. 

\section{Test procedure}

\begin{figure}
\begin{center}
  \includegraphics[width=.7\linewidth]{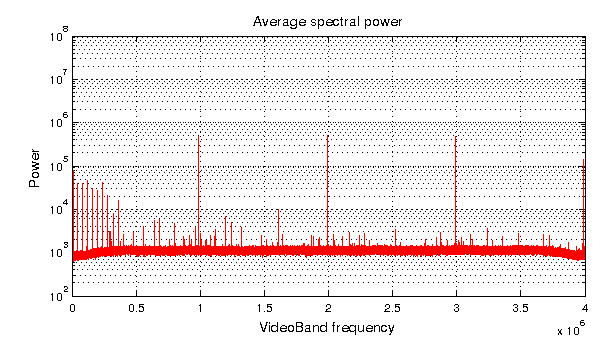}
\end{center}
\caption{A spectra with phase calibration spikes at 1 MHz intervals (at 1, 2, 3 and 4 MHz) from SWSpec.}
\label{fig:sws}
\end{figure}

\begin{figure}
\begin{center}
  \includegraphics[width=0.7\linewidth]{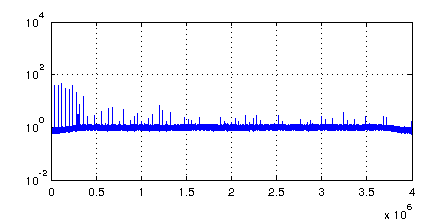}
\end{center}
\caption{A spectra without phase calibration from SWSpec.}
\label{fig:sws2}
\end{figure}

First the RF signal, which is locked to the 5 MHz reference, is fed to the receiver. Then a test measurement with Mark5 (using e.g. a VEX experiment procedure file in Field System to setup the DBBC correctly) is recorded in order to see how stable the RF signal's phase is when it is compared to the 5 MHz reference. The test data needs to be copied to a computer with SWSpec installed, and the scan name of the data recorded needs to be changed in the configuration file called inifile.Mh. The BandwidthHz parameter in the configuration file needs to have the same value as the bandwidth used in the DBBC channels during recording of the test data. After these run the SWSpec using the inifile and the data recorded, and plot the results with sctracking.m Matlab script. You need to select the correct input file, created by SWSpec, check that bandwidth is the same as during recording, and Nfft value is the same as inifile's FFTpoints value. Repeat the test, e.g., remove the phase reference, and look at the difference in the plots, see Figures \ref{fig:sws} and \ref{fig:sws2}.

\section{Conclusions}

Using SWSpec for VLBI signal chain testing when a problem occurs during an EVN session is quick and easy, and does not require any changes in the hardware setup. By using SWSpec instead of other measurement devices at least one hour is saved in the pre-operations per a VLBI session. SWSpec is also easier to use than the earlier hardware which you needed to gather, couple and setup correctly. In the future testing can be performed by operators that reduces workload of technical staff.

\pagebreak

\end{document}